\documentclass[showpacs,superscriptaddress,twocolumn,amsmath]{revtex4}
\usepackage{amssymb}
\usepackage{graphicx,color}
\usepackage{bm}
\usepackage[dvipdfm,bookmarks=true]{hyperref}

\newcommand{\ti}{\Tilde}

\newcommand{\nl}{\nonumber \\}
\newcommand{\nla}{\nl&\quad}

\newcommand{\up}{\uparrow}
\newcommand{\down}{\downarrow}

\newcommand{\be}{\begin{equation}}
\newcommand{\ee}{\end{equation}}
\newcommand{\bea}{\begin{eqnarray}}
\newcommand{\eea}{\end{eqnarray}}
\newcommand{\bsube}{\begin{subequations}}
\newcommand{\esube}{\end{subequations}}
\newcommand{\Eq}[1]{Eq.\,(\ref{#1})}

\newcommand{\dg}{\dagger}
\newcommand{\la}{\langle}
\newcommand{\ra}{\rangle}

\begin{document}
\draft

\topmargin=-40pt

\title{ Number-resolved master equation approach to quantum transport
            under the self-consistent Born approximation }

\author{ Yu Liu}
\affiliation{State Key Laboratory for Superlattices and Microstructures,
         Institute of Semiconductors,
         Chinese Academy of Sciences, Beijing 100083, China}
\author{ Jinshuang Jin}
\email{jsjin@hznu.edu.cn}
\affiliation{Department of Physics, Hangzhou Normal University,
          Hangzhou 310036, China}
\author{Jun Li}
\affiliation{ Beijing Computational Science Research Center,
              Beijing 100084, China }
\author{ Xin-Qi Li}
\email{lixinqi@bnu.edu.cn}
\affiliation{Department of Physics, Beijing Normal University,
          Beijing 100875, China}
\affiliation{State Key Laboratory for Superlattices and Microstructures,
         Institute of Semiconductors,
         Chinese Academy of Sciences, Beijing 100083, China}
\author{YiJing Yan}
\affiliation{Department of Chemistry, Hong Kong University of Science
             and Technology, Kowloon, Hong Kong}

\begin{abstract}
We construct a particle-number($n$)-resolved master equation (ME)
approach under the self-consistent Born approximation (SCBA)
for quantum transport through mesoscopic systems.
The formulation is essentially non-Markovian
and incorporates the interlay
of the multi-tunneling processes and many-body correlations.
The proposed $n$-SCBA-ME goes completely beyond the scope
of the Born-Markov master equation,
being applicable to transport under small bias voltage,
in non-Markovian regime and with strong Coulomb correlations.
For steady state, it can recover not only
the exact result of noninteracting transport under arbitrary voltages,
but also the challenging nonequilibrium Kondo effect.
Moreover, the $n$-SCBA-ME approach is efficient
for the study of shot noise.
We demonstrate the application by a couple of representative examples,
including particularly the nonequilibrium Kondo system.\\
\\
{\bf Master equation, quantum transport, shot noise spectrum}
\end{abstract}

\date{\today}
\pacs{73.23.-b,73.63.-b,72.10.Bg,72.90.+y}
\maketitle

\section{Introduction}

In addition to the Landauer-B\"uttiker scattering theory
and the non-equilibrium Green's function
method \cite{Dat95,Hau96}, as an alternative choice,
the rate or master equation approach is very convenient
for transport though nanostructures with a few discrete states
\cite{Gla88,Dav93,Naz93,Gur96a,Gur96b,Sch94,Sch96,Sch06,Li05a,Li05b}.
Moreover, the number($n$)-resolved version of the transport
master equation \cite{Gur96a,Gur96b,Li05a,Li05b,Shn98+01},
has been demonstrated as an efficient scheme for studies of
shot noise and counting statistics in mesoscopic transports,
including also the {\it large-derivation} analysis \cite{Li11}.

However, the perturbative master equation is usually up
to the 2nd-order expansion of the tunneling Hamiltonian,
which makes it applicable only in the limit of large bias voltage.
This 2nd-order master equation (2nd-ME) does not account for
the level's broadening effect.
Moreover, if applying to Coulomb interacting system, it cannot
describe cotunneling process and the nonequilibrium Kondo effect.
Therefore, higher-order expansions of the tunneling Hamiltonian
are necessary sometimes, as the efforts made in literature
\cite{Sch94,Sch96,Sch06,Yan080911,Wac05+10}.

In this work, by an insight from the Green's function theory,
we generalize the master equation approach
from the usual 2nd-order Born approximation (BA)
to self-consistent Born approximation (SCBA).
We will demonstrate that the effect of this improvement is remarkable:
it can recover not only the {\it exact} result
of noninteracting transport under arbitrary voltages, but also
the nonequilibrium Kondo effect in Coulomb interacting system.
In particular, the particle-number($n$)-resolved version of the SCBA-ME
($n$-SCBA-ME) provides an efficient scheme for studying the shot noise
and counting statistics, as to be illustrated by a couple of
application examples in this work.

The paper is organized as follows.
In Sec.\ II we outline the main formulation of the SCBA-ME,
where the steady state current and an illustrative example
will be presented.
In Sec.\ III, we continue the formal construction of the $n$-SCBA-ME
and provide the calculation scheme of noise spectrum,
while leaving the specific examples in Sec.\ IV
with in particular the shot noise of the nonequilibrium Kondo system.
Finally, we summarize the work in Sec.\ V.

\section{Formulation of the SCBA-ME}


In general we describe a transport setup by
$ H = H_S(a_{\mu}^{\dg},a_{\mu})+ H_{\rm res} + H'$.
Here $H_S$ is the Hamiltonian of the central
{\it system} embedded between two leads,
with $a^{\dg}_{\mu}$ ($a_{\mu}$) the creation
(annihilation) operator of the state $|\mu\ra$.
The other two Hamiltonians,
$H_{\rm res}$ and $H'$, describe the leads and their tunnel coupling
to the central system.
They are modeled by, respectively,
$H_{\rm res}=\sum_{\alpha=L,R}\sum_{k}\epsilon_{\alpha k}
b^{\dg}_{\alpha k}b_{\alpha k}$ and
$H'=\sum_{\alpha=L,R}\sum_{\mu k}(t_{\alpha\mu k}
a^{\dg}_{\mu}b_{\alpha k}+\rm{H.c.})$
with $b^{\dg}_{\alpha k}$ ($b_{\alpha k}$)
the creation (annihilation) operator of
electron in state $|k\ra$
of the left ($L$) and right ($R$) leads.

\subsection{ME under Born Approximation}

In a compact form, the master equation under the Born approximation
can be expressed as \cite{Li05b}
\begin{align}\label{BA-ME}
  \dot\rho(t) &=-i{\cal L}\rho(t)
  - \sum_{\mu\sigma}\Big\{\big[a^{\bar\sigma}_\mu,
  A^{(\sigma)}_{\mu\rho}(t)\big]
  +{\rm H.c.}      \Big\} .
\end{align}
In this work we use a reduced system of units
by setting $\hbar=k_B=e=1$ for the Planck constant,
the Boltzmann constant and the electron charge.
In \Eq{BA-ME} we also define:
$\sigma=+$ and $-$, $\bar{\sigma}=-\sigma$;
$a^+_{\mu}=a^{\dagger}_{\mu}$, and $a^-_{\mu}=a_{\mu}$.
The superoperators read
${\cal L}\rho=[H_S,\rho]$, and
$ A^{(\sigma)}_{\mu\rho}(t)
= \sum_{\alpha=L,R} A^{(\sigma)}_{\alpha\mu\rho}(t)$
while
$ A^{(\sigma)}_{\alpha\mu\rho}(t)
 =\sum_\nu\int^t_0 d\tau C^{(\sigma)}_{\alpha\mu\nu}(t-\tau)
\left\{{\cal G}(t,\tau)[a^{\sigma}_\nu\rho(\tau)]\right\} $.
${\cal G}(t,\tau)$ is the free propagator, determined by the
{\it system} Hamiltonian as
${\cal G}(t,\tau)=e^{-i{\cal L}(t-\tau)}$.

For the convenience of later use, we present a specific characterization for
$ C^{(\sigma)}_{\alpha\mu\nu}(t-\tau)$,
the correlation function of the reservoir electrons (in local equilibrium):
\begin{align}
C^{(\sigma)}_{\alpha\mu\nu}(t-\tau)
= \la f^{(\sigma)}_{\alpha\mu}(t)
f^{(\bar\sigma)}_{\alpha \nu}(\tau) \ra_{\rm B}.
\end{align}
Here, $f^{(+)}_{\alpha\mu}(t)=f^{\dagger}_{\alpha\mu}(t)$
and $f^{(-)}_{\alpha\mu}(t)=f_{\alpha\mu}(t)$,
via rewriting the tunneling Hamiltonian as
$H' = \sum_{\alpha=L,R}\sum_{\mu} \left( a^{\dg}_{\mu} f_{\alpha\mu}
       + \rm{H.c.}\right) $
by introducing
$f_{\alpha\mu} = \sum_{k} t_{\alpha\mu k}b_{\alpha\mu k}$.
The time dependence of the operators in
$ C^{(\sigma)}_{\alpha\mu\nu}(t-\tau)$
originates from the interaction picture
with respect to the reservoir Hamiltonian,
and the average $\la \cdots \ra_B$ is over the reservoir states.
Moreover, we introduce the Fourier transform of
$ C^{(\sigma)}_{\alpha\mu\nu}(t-\tau)$:
\begin{align}
C^{(\pm)}_{\alpha\mu\nu}(t-\tau)=\int^\infty_{-\infty}
\frac{d\omega}{2\pi}
e^{\pm i\omega (t-\tau)}\Gamma^{(\pm)}_{\alpha\mu\nu}(\omega).
\end{align}
Accordingly, we have
$\Gamma^{(+)}_{\alpha\mu\nu}(\omega)
=\Gamma_{\alpha\nu\mu}(\omega)n^{(+)}_{\alpha}(\omega)$
and $\Gamma^{(-)}_{\alpha\mu\nu}(\omega)
=\Gamma_{\alpha\mu\nu}(\omega)n^{(-)}_{\alpha}(\omega)$,
where $\Gamma_{\alpha\mu\nu}(\omega)
=2\pi\sum_{k}t_{\alpha\mu k}t^\ast_{\alpha\nu k}\delta(\omega-\epsilon_k)$
is the spectral density function of the reservoir ($\alpha$),
$n^{(+)}_{\alpha}(\omega)$ denotes the Fermi function $n_{\alpha}(\omega)$,
and $n^{(-)}_{\alpha}(\omega)=1-n_{\alpha}(\omega)$ is introduced for brevity.
Alternatively, we may introduce as well the Laplace transform
of $ C^{(\sigma)}_{\alpha\mu\nu}(t-\tau)$, denoting by
$C^{(\sigma)}_{\alpha\mu \nu}(\omega)$,
which is related with $\Gamma^{(\pm)}_{\alpha\mu \nu}(\omega)$
through the well known dispersive relation:
\begin{align}\label{FDT2}
C^{(\pm)}_{\alpha\mu \nu}(\omega)
&=\int^\infty_{-\infty}\frac{d\omega'}{2\pi}
\frac{i}{\omega\pm\omega'+i0^+}\Gamma^{(\pm)}_{\alpha\mu \nu}(\omega').
\end{align}

In this work, for the reservoir spectral density function,
we assume a Lorentzian form as
\be\label{Gammaw}
\Gamma_{\alpha\mu \nu}(\omega)=
\frac{\Gamma_{\alpha\mu \nu} W^2_\alpha}{(\omega-\mu_\alpha)^2+W^2_\alpha} .
\ee
In some sense, this assumption corresponds to a half-occupied band
for each lead, which peaks the Lorentzian center
at the chemical potential $\mu_\alpha$.
$W_\alpha$ characterizes the bandwidth of the $\alpha$th lead.
Obviously, the usual constant spectral density function is
recovered from \Eq{Gammaw} in the limit $W_\alpha\rightarrow\infty$,
yielding $\Gamma_{\alpha\mu \nu}(\omega)=\Gamma_{\alpha\mu \nu}$.
Corresponding to the above Lorentzian spectral density function,
straightforwardly, we obtain
\begin{align}
C^{(\pm)}_{\alpha\mu \nu}(\omega)
&=\frac{1}{2}\left[\Gamma^{(\pm)}_{\alpha\mu \nu}(\mp \omega)
 +i\Lambda^{(\pm)}_{\alpha\mu \nu}(\mp \omega)\right].
\end{align}
The imaginary part, through the dispersive relation,
is associated with the real one as
\begin{align}
& \Lambda^{(\pm)}_{\alpha\mu \nu} (\omega)
={\cal P}\int^\infty_{-\infty}\frac{d\omega'}{2\pi}
\frac{1}{\omega\pm\omega'}\Gamma^{(\pm)}_{\alpha\mu \nu}(\omega)
\nl&=\frac{\Gamma_{\alpha\mu\nu}}{\pi}
\Bigg\{{\rm Re}\left[\Psi\left(\frac{1}{2}
+i\frac{\beta(\omega-\mu_\alpha)}{2\pi}\right)\right]
\nla
-\Psi\left(\frac{1}{2}+\frac{\beta W_\alpha}{2\pi}\right)
\mp\pi\frac{\omega-\mu_\alpha}{W_\alpha}\Bigg\},
\end{align}
where ${\cal P}$ stands for the principle value
and $\Psi(x)$ is the digamma function.

We remark that the 2nd-order master equation can apply only to transport
under large bias voltage. That is, the Fermi levels of the leads
should be considerably away from the system levels,
by at least several times of the level's broadening.

\subsection{ME under Self-Consistent Born Approximation}

The basic idea to improve the 2nd-ME can follow what is typically done
in the Green's function theory, i.e., correcting the self-energy diagram
from the {\it Born} to a {\it self-consistent Born} approximation.
In our case, the SCBA scheme can be implemented by
replacing the {\it free} propagator
in the 2nd-order master equation,
${\cal G}(t,\tau)=e^{-i{\cal L}(t-\tau)}$,
by an effective one, ${\cal U}(t,\tau)$,
which propagates a state
with the precision of the 2nd-order Born approximation.
From this type of consideration, the generalized SCBA-ME
follows \Eq{BA-ME} directly as \cite{LJL11}:
\begin{align}\label{SCBA-ME}
  \dot\rho(t) &=-i{\cal L}\rho(t)
  - \sum_{\mu\sigma}\Big\{\big[a^{\bar\sigma}_\mu,
  {\cal A}^{(\sigma)}_{\mu\rho}(t)\big]
  +{\rm H.c.}      \Big\} .
\end{align}
Here $ {\cal A}^{(\sigma)}_{\mu\rho}(t)
= \sum_{\alpha=L,R} {\cal A}^{(\sigma)}_{\alpha\mu\rho}(t)$,
and $ {\cal A}^{(\sigma)}_{\alpha\mu\rho}(t)
 =\sum_\nu\int^t_0 d\tau C^{(\sigma)}_{\alpha\mu\nu}(t-\tau)
\left\{{\cal U}(t,\tau)[a^{\sigma}_\nu\rho(\tau)]\right\} $.
To close this master equation, let us define
$\ti\rho_j(t)\equiv {\cal U}(t,\tau)[a^{\sigma}_\nu\rho(\tau)]$
(here and in the following we use ``$j$" to denote the double
indices $(\nu,\sigma)$ for the sake of brevity).
Then, the equation-of-motion (EOM) of this auxiliary object reads
\begin{align}\label{tirhoj}
  \dot{\ti\rho}_j(t) = -i{\cal L}\ti\rho_j(t)
-\int^t_{\tau}dt'  \Sigma^{(A)}_2(t-t')\ti\rho_j(t').
\end{align}
In this equation the 2nd-order self-energy superoperator,
$\Sigma^{(A)}_2(t-t')$, differs from the usual one
because it involves {\it anticommutators}, rather than
the {\it commutators} in the 2nd-order master equation.
More explicitly, we have
\begin{align}\label{Acmt}
\int^t_{\tau}dt' \Sigma^{(A)}_2(t-t')
& \ti\rho_j(t')= \sum_{\mu} \Big[\big\{a_\mu,A^{(+ )}_{\mu\ti\rho_j}\big\}
+\big\{a^\dg_\mu,A^{(-)}_{\mu\ti\rho_j}\big\}   \nl
& +\big\{a^\dg_\mu,A^{(+ )\dg}_{\mu\ti\rho_j}\big\}
+\big\{a_\mu,A^{(- )\dg}_{\mu\ti\rho_j}  \big\} \Big] ,
\end{align}
where $A^{(\pm)}_{\mu\ti\rho_j}$ is defined as
$ A^{(\sigma')}_{\mu\ti\rho_j}=\sum_{\alpha=L,R}\sum_{\nu'}
 \int^t_\tau dt' C^{(\sigma')}_{\alpha\mu\nu'}(t-t')
\left\{ e^{-i{\cal L}(t-t')}[a^{\sigma'}_{\nu'}
\tilde{\rho}_j(t')]\right\}$.
Because of the {\it anticommutative} brackets here,
we stress that the propagation of $\ti\rho_j(t)$
does not satisfy the usual 2nd-order master equation.
This, in certain sense, violates the so-called {\it quantum regression theorem}.

\subsection{Steady State Current}

Within the framework of SCBA-ME,
similar to its 2nd-order counterpart,
current through the $\alpha$th lead reads
\begin{align}
   I_{\alpha}(t)= 2\sum_\mu {\rm Re}
\left\{ {\rm Tr}\big[ {\cal A}^{(+)}_{\alpha\mu\rho}(t)a_\mu
  - {\cal A}^{(-)}_{\alpha\mu\rho}(t)a^\dg_\mu\big] \right\} .
\end{align}
For steady state,
consider the integral $\int^t_0 d\tau [\cdots]\rho(\tau)$
in ${\cal A}^{(\pm)}_{\alpha\mu\rho}(t)$.
Since physically, the correlation function
$C^{(\pm)}_{\alpha\mu\nu}(t-\tau)$ in the integrand
is nonzero only on {\it finite} timescale,
we can replace $\rho(\tau)$
in the integrand by the steady state $\bar{\rho}$,
in the long time limit ($t\rightarrow\infty$).
After this replacement, we obtain
\begin{align}
{\cal A}^{(\pm)}_{\alpha\mu\bar\rho} 
=\sum_\nu\int^\infty_{-\infty}\frac{d\omega}{2\pi}\,
\Gamma^{(\pm)}_{\alpha\mu\nu}(\omega)
{\cal U}(\pm\omega)[a^{\pm}_\nu\bar\rho] .
\end{align}
Then, substituting this result into \Eq{SCBA-ME},
we can straightforwardly solve for $\bar\rho$
and calculate the steady state current.

Based on $\bar{\rho}$, to obtain further the current, we first introduce
$\varphi_{1\mu\nu}(\omega)={\rm Tr} \big[a_\mu\ti\rho_{1\nu}(\omega)\big]$
and
$\varphi_{2\mu\nu}(\omega)={\rm Tr} \big[a_\mu\ti\rho_{2\nu}(\omega)\big]$,
where $\ti\rho_{1\nu}(\omega)$ and $\ti\rho_{2\nu}(\omega)$
are calculated using \Eq{tirhoj},
with an initial condition of
$\ti\rho_{1\nu}(0)=\bar\rho a^\dg_\nu$
and $\ti\rho_{2\nu}(0)=a^\dg_\nu\bar\rho$.
To simplify notations, we denote the various matrices
in boldface form: $\bm\varphi_1(\omega)$,
$\bm\varphi_2(\omega)$ and $\bm{\Gamma}_{L(R)}$.
Now, if $\bm\Gamma_L$ is proportional to $\bm\Gamma_R$ by a constant,
the steady state current can be recast to the Landauer-B\"uttiker type:
\begin{align}
\bar I =  2~ {\rm Re}
\int^\infty_{-\infty}\frac{d\omega}{2\pi}
\left[ n_L(\omega)- n_R(\omega)\right] {\cal T}(\omega) ,
\end{align}
where tunneling coefficient, very compactly, is given by
\begin{align}
 {\cal T}(\omega)={\rm Tr}\{
  \bm\Gamma_L\bm\Gamma_R
  (\bm\Gamma_L+\bm\Gamma_R)^{-1}
  {\rm Re}\big[\bm\varphi(\omega)\big]\}.
\end{align}
Here $\bm\varphi(\omega)=\bm\varphi_1(\omega)+\bm\varphi_2(\omega)$.

Now we demonstrate that, for a {\it noninteracting system},
the above stationary current
coincides precisely with the nonequilibrium Green's function approach,
both giving the {\it exact} result under arbitrary bias voltage.
In general, a noninteracting system can be described by
$H_S=\sum_{\mu\nu}h_{\mu\nu}a^\dg_\mu a_\nu$. Straightforwardly,
we obtain the EOM for $\bm \varphi_i$ as follows:
\begin{align}\label{varphiw0}
-i\omega\bm\varphi_i(\omega)-\bm\varphi_i(0)=-i\bm h\bm\varphi_i(\omega)
-i\bm\Sigma_{0}(\omega)\bm\varphi_i(\omega).
\end{align}
$\bm\varphi_i(0)$ stand for the initial conditions,
$\varphi_{1\mu\nu}(0)={\rm Tr}\big[a_\mu\bar\rho a^\dg_{\nu}\big]$
and $\varphi_{2\mu\nu}(0)={\rm Tr}\big[a_\mu a^\dg_{\nu}\bar\rho\big]$.
The tunnel-coupling self-energy $\bm\Sigma_{0}$ reads
$ \Sigma_{0\mu\nu}(\omega)
= -i\sum_{\alpha} \big[ C^{(-)}_{\alpha\mu\nu}(\omega)
+ C^{(+)\ast}_{\alpha\mu\nu}(-\omega)\big]$, or
\begin{align}\label{Self1}
\Sigma_{0\mu\nu}(\omega)
&=\int^\infty_{-\infty} \frac{d\omega'}{2\pi}
\frac{\Gamma_{\mu\nu}(\omega')}{\omega-\omega' +i0^+} .
\end{align}
Then, based on \Eq{varphiw0}, summing up $\bm\varphi_1(\omega)$
and $\bm\varphi_2(\omega)$ yields
\be\label{phiomega}
\bm\varphi(\omega)
=i\big[ \omega-\bm h-\bm\Sigma_{0}(\omega)\big]^{-1}
\ee
In deriving this result, the cyclic property under trace
and the anti-commutator,
$\{a_\mu,a^\dg_\nu\}=\delta_{\mu\nu}$, have been used.
\Eq{phiomega} is nothing but the {\it exact} Green's function
for transport through a noninteracting system,
giving thus the exact stationary current after inserting it
into the above current formula.

\subsection{Interacting Case}

To show the application of the proposed SCBA-ME
to {\it interacting system}, as an illustrative example,
we consider the transport through
an interacting quantum dot described as
\be\label{Ands-H}
H_S = \sum_{\mu}\left(\epsilon_{\mu} a_{\mu}^{\dg}a_{\mu}
       +\frac{U}{2}n_{\mu}n_{\bar{\mu}}\right) .
\ee
Here the index $\mu$ labels the spin up (``$\uparrow$") and spin
down (``$\downarrow$") states, and $\bar{\mu}$ stands for the
opposite spin orientation.
$\epsilon_{\mu}$ denotes the spin-dependent energy level,
which may account for the Zeeman splitting in the presence of
magnetic field ($B$),
$\epsilon_{\uparrow,\downarrow}=\epsilon_0\pm g\mu_B B$.
Here $\epsilon_0$ is the degenerate dot level
in the absence of magnetic field;
$g$ and $\mu_B$ are, respectively, the Lande-$g$ factor and the Bohr's magneton.
In the interaction part, say, the Hubbard term
$Un_{\uparrow}n_{\downarrow}$,
$n_{\mu}=a^{\dg}_{\mu}a_{\mu}$ is the number operator
and $U$ represents the interacting strength.

First, we note that $C^{(\pm)}_{\alpha\mu\nu}$
is diagonal with respect to the spin states, i.e.,
$C^{(\pm)}_{\alpha\mu\nu}(t)=\delta_{\mu\nu}C^{(\pm)}_{\alpha\mu}(t)$
and
$\Gamma^{(\pm)}_{\alpha\mu\nu}=\Gamma^{(\pm)}_{\alpha\mu}\delta_{\mu\nu}$.
Then, we specify the states involved in the transport as
$|0\ra$, $|\up\ra$, $|\down\ra$ and $|d\ra$, corresponding to
the empty, spin-up, spin-down and double occupancy states, respectively,
Using this basis, we reexpress the electron operator
in terms of projection operator,
$a^\dg_\mu=|\mu\ra\la 0|+(-1)^{\mu}|d\ra\la \bar \mu|$,
where the convention $(-1)^\up=1$ and $(-1)^\down=-1$ is implied.
For a solution of the steady state, we have
\begin{align}\label{Arhost-Ad}
{\cal A}^{(\pm)}_{\alpha\mu\bar\rho}
&=\int^\infty_{-\infty}\frac{d\omega}{2\pi}\,
\Gamma^{(\pm)}_{\alpha\mu}(\omega)
{\cal U}(\pm\omega)[a^{\pm}_\mu\bar\rho].
\end{align}
Straightforwardly, after some algebra,
${\cal U}(\pm\omega)[a^{\pm}_\mu\bar\rho]$
can be carried out as
\begin{align}
{\cal U}(\omega)[a^{\dg}_\mu\bar\rho]
&=\left[\lambda^+_\mu(\omega)|\mu\ra\la 0|
+\kappa^+_\mu(\omega)(-1)^{\mu}|d\ra\la \bar \mu|\right] , \nonumber
\\
{\cal U}(-\omega)[a_\mu\bar\rho]&=\left[\lambda^-_\mu(\omega)|0\ra\la \mu|
+\kappa^-_\mu(\omega)(-1)^{\mu}|\bar \mu\ra\la d|\right] ,
\end{align}
where
\begin{align}
\lambda^+_\mu(\omega)&=i\frac{ \Pi^{-1}_{1\mu}(\omega)\bar\rho_{00}
-\Sigma^-_{\bar\mu}(\omega)\bar\rho_{\bar\mu\bar\mu}}
{ \Pi^{-1}_{\mu}(\omega)\Pi^{-1}_{1\mu}(\omega)},  \nonumber
\\
\lambda^-_\mu(\omega)&=i\frac{ \Pi^{-1}_{1\mu}(\omega)\bar\rho_{\mu\mu}
-\Sigma^-_{\bar\mu}(\omega)\bar\rho_{dd} }
{ \Pi^{-1}_{\mu}(\omega)\Pi^{-1}_{1\mu}(\omega)}, \nonumber
\\
\kappa^+_\mu(\omega)&=i\frac{ -\Sigma^+_{\bar\mu}(\omega)\bar\rho_{00}
+\Pi^{-1}_{\mu}(\omega)\bar\rho_{\bar\mu\bar\mu} }
{ \Pi^{-1}_{\mu}(\omega)\Pi^{-1}_{1\mu}(\omega)},  \nonumber
\\
\kappa^-_\mu(\omega)&=i\frac{ -\Sigma^+_{\bar\mu}(\omega)\bar\rho_{\mu\mu}
+\Pi^{-1}_{\mu}(\omega)\bar\rho_{dd} }
{ \Pi^{-1}_{\mu}(\omega)\Pi^{-1}_{1\mu}(\omega)}. \nonumber
\end{align}
Here, we introduced
$\Pi^{-1}_{ \mu}(\omega)=\omega-\epsilon_\mu-\Sigma_{0\mu}(\omega)
 -\Sigma^+_{\bar\mu}(\omega)$, and
$\Pi^{-1}_{1\mu}(\omega)=\omega-\epsilon_\mu-U-\Sigma_{0\mu}(\omega)
-\Sigma^-_{\bar\mu}(\omega)$.
The self-energies $\Sigma_{0\mu}(\omega)$ and $\Sigma^{\pm}_{\mu}(\omega)$
are given by
\begin{align}
\Sigma_{0\mu}(\omega)
&=\int^\infty_{-\infty} \frac{d\omega'}{2\pi}
\frac{\Gamma_{\mu}(\omega')}{\omega-\omega' +i0^+} , \nl
\Sigma^{\pm}_{\mu}(\omega)
&=\int^\infty_{-\infty} \frac{d\omega'}{2\pi}
\frac{\Gamma^{(\pm)}_{\mu}(\omega')}{\omega-\epsilon_{\bar\mu}
+\epsilon_\mu-\omega' +i0^+}
\nl&
+\int^\infty_{-\infty} \frac{d\omega'}{2\pi}
\frac{\Gamma^{(\pm)}_{\mu}(\omega')}{\omega-E_d+\omega' +i0^+}.
\end{align}
Then, we find the solution of $\bm\varphi(\omega)$ as
\begin{align}\label{Kondo}
& \bm\varphi(\omega)
=
\frac{i(1-n_{\bar\mu})}
{\omega-\epsilon_\mu-\Sigma_{0\mu}
 +U\Sigma^+_{\bar\mu}(\omega-\epsilon_\mu-U-\Sigma_{0\mu}
 -\Sigma_{\bar\mu})^{-1}
 }   \nl
& ~
+\frac{ i n_{\bar\mu}}
{\omega-\epsilon_\mu-U-\Sigma_{0\mu}
 -U\Sigma^-_{\bar\mu}(\omega-\epsilon_\mu-\Sigma_{0\mu}
 -\Sigma_{\bar\mu})^{-1}
 },
\end{align}
where $n_\mu=\rho_{\mu\mu}+\rho_{dd}$, and
$1-n_\mu=\rho_{\bar\mu\bar\mu}+\rho_{00}$.
This result, precisely, coincides with the one from the EOM
technique of the nonequilibrium Green's function (nGF) \cite{Hau96}.
Therefore, as discussed in detail in the book by Haug and Jauho \cite{Hau96},
this solution contains the remarkable nonequilibrium Kondo effect.

At high temperatures, the terms $U\Sigma^{\pm}_{\bar\mu}(\cdots)^{-1}$
vanish, reducing thus \Eq{Kondo} to
\begin{align}\label{HF}
\bm\varphi_{HF}(\omega) = \frac{i(1-n_{\bar\mu})}
{\omega-\epsilon_\mu-\Sigma_{0\mu} }
+  \frac{ i n_{\bar\mu}}
{\omega-\epsilon_\mu-U-\Sigma_{0\mu}  }.
\end{align}
Here we use $\bm\varphi_{HF}$ to imply the result
at the level of a mean-field Hatree-Fock approximation.
Actually, \Eq{HF} can also be derived from the EOM technique
of nGF at lower-order cutoff, by using a mean-field approximation \cite{Hau96}.
The point is that, noting the {\it broadening effect} contained,
even this simple result goes beyond the scope of
the 2nd-order master equation.
In Fig.\ 1 we plot the current-voltage relation
based on \Eq{Kondo} against that from \Eq{HF}.

\begin{figure}[h]
\includegraphics[width=6cm]{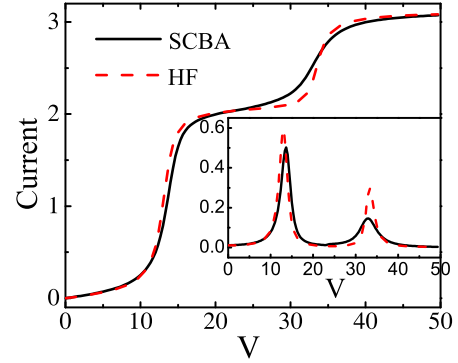}
\caption{
Coulomb staircase in the current-voltage curve.
Inset: the corresponding differential conductance.
The result based on \Eq{Kondo} is plotted
against the Hatree-Fock (HF) solution \Eq{HF}.
Parameters: $\Gamma_L=\Gamma_R=\Gamma/2$,
$\epsilon_0=7\Gamma$, $U=10\Gamma$, and $k_BT=0.1\Gamma$.
The bias voltage is set as $\mu_L=-\mu_R=eV/2$ which
assumes the zero-bias Fermi level as energy reference.
In this work (here and in other figures below),
we use a reduced system of units by assuming $\hbar=e=k_B=1$,
and setting $\Gamma=1$ for an arbitrary unit of energy.  }
\end{figure}

\section{Formulation of the $n$-SCBA-ME}

Now we proceed to construct the particle number (``$n$")
resolved SCBA-ME, along the same line in constructing the
``$n$"-resolved 2nd-order master equation \cite{Li05a,Li05b}.
The basic idea is to split the Hilbert space of the reservoirs
into a set of subspaces, each labeled by $n$.
Then, do the average (trace) over each subspace
and define the corresponding reduced density matrix as $\rho^{(n)}(t)$.
To be specific, consider the $\rho^{(n)}(t)$
conditioned on the electron number
arrived to the {\it right} lead, which obeys
\begin{align}\label{nME-scba}
\dot{\rho}^{(n)}
   & =  -i {\cal L}\rho^{(n)} -  \sum_{\mu}
   \Big\{\big [a_{\mu}^{\dg} {\cal A}_{\mu\ti{\rho}^{(n)}}^{(-)}
   +a_{\mu} {\cal A}_{\mu\ti{\rho}^{(n)}}^{(+)}
    -  {\cal A}_{L\mu\ti{\rho}^{(n)}}^{(-)}a_{\mu}^{\dg}
  \nl
&  -  {\cal A}_{L\mu\ti{\rho}^{(n)}}^{(+)}a_{\mu}
   -  {\cal A}_{R\mu\ti{\rho}^{(n-1)}}^{(-)}a_{\mu}^{\dg}
    -  {\cal A}_{R\mu\ti{\rho}^{(n+1)}}^{(+)}a_{\mu}\big]
 +{\rm H.c.} \Big\}  .
\end{align}
Here $ {\cal A}^{(\sigma)}_{\alpha\mu\ti{\rho}^{(n)}}(t)
=\sum_\nu\int^t_0 d\tau C^{(\sigma)}_{\alpha\mu\nu}(t-\tau)
[ \ti{\rho}_j^{(n)}(t,\tau)] $, while the summation over $\nu$
makes sense in regard to the abbreviation $j=\{\nu,\sigma\}$.
In \Eq{nME-scba},
the appearing of $\tilde{\rho}^{(n\pm 1)}_j(t,\tau)$
is owing to a more tunneling event (forward/backword)
involved in the process of the corresponding terms.
In particular, $\ti{\rho}_j^{(n)}(t,\tau)$
is the $n$-dependent version of the quantity
$\ti{\rho}_j(t,\tau)={\cal U}(t,\tau)[a^{\sigma}_{\nu}\rho(\tau)]$,
satisfying an EOM according to \Eq{tirhoj}:
\begin{align}\label{nME-2nd}
\dot{\ti\rho}^{(n)}_{j}
   &=  -i {\cal L}\ti\rho^{(n)}_{j} -  \sum_{\mu}
   \Big\{\big [a_{\mu}^{\dg} A_{\mu\tilde{\rho}^{(n)}_{j}}^{(-)}
   +a_{\mu} A_{\mu\tilde{\rho}^{(n)}_{j}}^{(+)}
    +  A_{L\mu\tilde{\rho}^{(n)}_{j}}^{(-)}a_{\mu}^{\dg}
\nl& 
   +  A_{L\mu\tilde{\rho}^{(n)}_{j}}^{(+)}a_{\mu}
   +  A_{R\mu\tilde{\rho}^{(n-1)}_{j}}^{(-)}a_{\mu}^{\dg}
    +  A_{R\mu\tilde{\rho}^{(n+1)}_{j}}^{(+)}a_{\mu}\big]
 +{\rm H.c.} \Big\} .
\end{align}
In this equation we introduced
$ A^{(\sigma')}_{\alpha\mu\ti\rho^{(n)}_{j}}(t)
 =\sum_{\nu'}\int^t_\tau dt' C^{(\sigma')}_{\alpha\mu\nu'}(t-t')
\left\{ e^{-i{\cal L}(t-t')}[a^{\sigma'}_{\nu'}
\tilde{\rho}_j^{(n)}(t')]\right\}$.


\vspace{0.2cm}

The $n$-resolved master equation contains rich information.
This allows a great variety of its applications, including
such as a convenient calculation of shot noise spectrum
and the study of full counting statistics \cite{Li11}.
In the remaining part of this work we focus on
the issue of shot noise spectrum.
The noise spectrum, $S(\omega)$, is the Fourier transform
of the current correlation function
$S(t)=\la I(t) I(0)\ra_{ss}$ defined in the steady state.
Very conveniently, within the framework of the $n$-ME,
one can calculate $S(\omega)$
by using the MacDonald's formula \cite{Li05a}:
$S(\omega)=2\omega\int^{\infty}_{0}dt \sin(\omega t)
\frac{d}{dt}\langle n^{2}(t)\rangle$,
where
$\langle n^{2}(t)\rangle = \sum_{n}n^{2}P(n,t)= {\rm Tr}
\sum_{n} n^{2} \rho^{(n)}(t)$, and the $n$-counting
starts with the steady state ($\bar{\rho}$).
Based on \Eq{nME-scba}, one can express
$\frac{d}{dt} \langle n^{2}(t)\rangle$
in terms of ${\cal A}^{(\sigma)}_{R\mu\bar{\rho}}(t)$
and ${\cal A}^{(\sigma)}_{R\mu\ti{N}}(t)$.
The former has been introduced in \Eq{SCBA-ME},
needing only to replace $\rho(\tau)$ by $\bar{\rho}$.
The latter reads $ {\cal A}^{(\sigma)}_{R\mu\ti{N}}(t)
=\sum_\nu\int^t_0 d\tau C^{(\sigma)}_{R\mu\nu}(t-\tau)
[ \ti{N}_j(t,\tau)]$, where
$\ti{N}_j(t,\tau)=\sum_n n \ti{\rho}_j^{(n)}(t,\tau)$.
Then, the MacDonald's formula becomes
\begin{align}\label{Sw-scba}
S(\omega)&= 2\omega {\rm Im}\sum_{\mu}{\rm Tr}\Big\{ 2 \big[
 {\cal A}_{R\mu \ti{N}}^{(-)}(\omega)a_{\mu}^{\dg}
    - {\cal A}_{R\mu \ti{N}}^{(+)}(\omega)a_{\mu}\big]
 \nl&\qquad\qquad
  +  \big[
  {\cal A}_{R\mu\bar{\rho}}^{(-)}(\omega)a_{\mu}^{\dg}
    + {\cal A}_{R\mu\bar{\rho}}^{(+)}(\omega)a_{\mu}\big]
  \Big\} .
\end{align}
This result is obtained after Laplace transforming
${\cal A}^{(\sigma)}_{R\mu\bar{\rho}}(t)$
and ${\cal A}^{(\sigma)}_{R\mu\ti{N}}(t)$. More explicitly,
\be
{\cal A}^{(\sigma)}_{R\mu\bar{\rho}}(\omega)
= \sum_\nu\int^\infty_{-\infty} \frac{d\omega'}{2\pi}
\Gamma^{(\sigma)}_{R\mu\nu}(\omega')
{\cal U}(\omega+\sigma\omega')[a^{\sigma}_\nu\bar{\rho}(\omega)], \nonumber
\ee
where the Laplace transformation of the steady state
reads $\bar{\rho}(\omega)=i\bar{\rho}/\omega$,
and the propagator ${\cal U}$ in frequency domain
is defined through \Eq{tirhoj}.
On the other hand, ${\cal A}^{(\sigma)}_{R\mu \ti{N}}(\omega)$ reads
\be
{\cal A}^{(\sigma)}_{R\mu \ti{N}}(\omega)
= \sum_\nu\int^\infty_{-\infty} \frac{d\omega'}{2\pi}
\Gamma^{(\sigma)}_{R\mu\nu}(\omega')
\ti{\cal U}(\omega+\sigma\omega')[a^{\sigma}_{\nu}N(\omega)].  \nonumber
\ee
In deriving this result, we introduced an additional propagator
through $\ti{N}_j(t,\tau)=\ti{\cal U}(t-\tau)\ti{N}_j(\tau)$,
with $\ti{N}_{j}(\tau)=a^{\sigma}_{\nu} N(\tau)$ as the initial condition
which is defined by $N(\tau)=\sum_n n \rho^{(n)}(\tau)$.
$\ti{\cal U}(\omega)$ and $N(\omega)$ can be obtained via
Laplace transforming the following EOMs.
{\it (i)} For $N(\omega)$,
based on the $n$-SCBA-ME we obtain
\begin{align}\label{N-t}
  \dot{N}(t) &=-i{\cal L}N(t)
  - \sum_{\mu\sigma}\Big\{\big[a^{\bar\sigma}_\mu,
  {\cal A}^{(\sigma)}_{\mu N}(t)\big]
  +{\rm H.c.}  \Big\}  \nl
& ~~ + \sum_{\mu} \Big\{\big[ {\cal A}_{R\mu\bar{\rho}}^{(-)}a_{\mu}^{\dg}
    - {\cal A}_{R\mu\bar{\rho}}^{(+)}a_{\mu}\big] +{\rm H.c.} \Big\}  .
\end{align}
{\it (ii)}
For $\ti{\cal U}(\omega)$, from \Eq{nME-2nd} we have
\begin{align}\label{Nj-2nd}
& \dot{\ti N}_{j}(t)
   =  -i {\cal L}\ti N_{j}(t)
   - \int^t_{\tau} dt' {\Sigma}^{(A)}_2(t-t')\ti N_{j}(t')
\nl& ~
 - \sum_{\mu} \Big\{\big [A_{R\mu\ti{\rho}_{j}}^{(-)}(t)a_{\mu}^{\dg}
-A_{R\mu\ti{\rho}_{j}}^{(+)} (t)a_{\mu}\big]
 +{\rm H.c.} \Big\}.
\end{align}
The self-energy superoperator ${\Sigma}^{(A)}_2(t-t')$
is referred to \Eq{Acmt} for its definition.
Similar as introduced in \Eq{nME-2nd}, we defined here
$ A^{(\sigma')}_{R\mu\ti\rho_{j}}(t)
 =\sum_{\nu'}\int^t_\tau dt' C^{(\sigma')}_{R\mu\nu'}(t-t')
\left\{ e^{-i{\cal L}(t-t')}[a^{\sigma'}_{\nu'}
\tilde{\rho}_j(t')]\right\}$.

For the convenience of application, we summarize
the solving protocol in a more transparent way as follows.
First, solve ${\cal U}(\omega)$ from \Eq{tirhoj}
and obtain $\rho(\omega)$ from \Eq{SCBA-ME};
then, extract $\tilde{\cal U}(\omega)$ from \Eq{Nj-2nd}
and $N(\omega)$ from \Eq{N-t}.
With the help of ${\cal U}(\omega)$,
$\tilde{\cal U}(\omega)$ and $N(\omega)$, one can
straightforwardly calculate the noise spectrum of \Eq{Sw-scba}.

%

\section{Illustrative Applications}

\subsection{Noninteracting Quantum Dot }

We consider the simplest case of transport
through a single-level quantum dot.
In the absence of magnetic field and Coulomb interaction,
the spin is an irrelevant degree of freedom
which is thus neglected in this example.
Then, the system Hamiltonian reads $H_S = \epsilon_0 a^{\dg}a$,
and the states involved in the transport are $|0\ra$ and $|1\ra$,
corresponding to the empty and occupied dot states.
Along the solving protocol outlined above,
it is straightforward to obtain the shot noise spectrum
as shown in Fig.\ 2 by the solid curve.
Shown also there, by the dashed and dotted curves, are the results
from the 2nd-order non-Markovian and Markovian master equation
(2nd-nMKV/MKV-ME). The former is based on Ref.\ \cite{Jin2011},
while the later is from the following analytic result \cite{Luo2007}:
\begin{align}
S(\omega)=2\bar{I} \left(
\frac{\Gamma^2_L+\Gamma^2_R+\omega^2}{\Gamma^2+\omega^2} \right),
\end{align}
where $\Gamma=\Gamma_L+\Gamma_R$ is assumed.
$\bar{I}$ is the steady state current, in large bias limit which
simply reads $\bar{I}=\Gamma_L\Gamma_R/\Gamma$, while here
we account for the finite bias effect based on the SCBA-ME approach.

We observe that, quantitatively, the result from the $n$-SCBA-ME
modifies that from the 2nd-nMKV-ME, while qualitatively both revealing
a {\it staircase} behavior at frequency
around $\omega_{\alpha 0}=|\mu_{\alpha}-\epsilon_0|$.
Mathematically, the origin of the staircase is from
the {\it time-nonlocal memory effect}.
Physically, this behavior is owing to the detection-energy ($\omega$)
assisted transmission resonance between the dot and leads,
which experiences a sharp change when crossing the Fermi levels.
In high frequency regime, the noise spectrum from the $n$-SCBA-ME
coincides with that from the 2nd-nMKV-ME, while the latter
is given in Ref.\ \cite{Jin2011} by the high frequency limit
as $S(\omega\rightarrow\infty)=\Gamma_R$.
This, straightforwardly, leads to a Fano factor
as $F=S/2\bar{I}=(1+\Gamma_R/\Gamma_L)/2$.
Therefore, it can be Poissonian, sub-Poissonian,
and super-Poissonian, depending on the symmetry factor $\Gamma_R/\Gamma_L$.
In contrast, the 2nd-nMKV-ME predicts
a Poissonian result, $F(\omega\rightarrow\infty)=1$.

We would like to remark that the 2nd-MKV-ME is only applicable
in the low frequency regime of
$\omega<\omega_{\alpha 0}=|\mu_{\alpha}-\epsilon_0|$.
This is in consistency with the fact that
the high frequency regime corresponds to a short
timescale where the non-Markovian effect is strong,
while the low frequency regime corresponds to a
long timescale where the non-Markovian effect diminishes.

\begin{figure}
\includegraphics[scale=0.25]{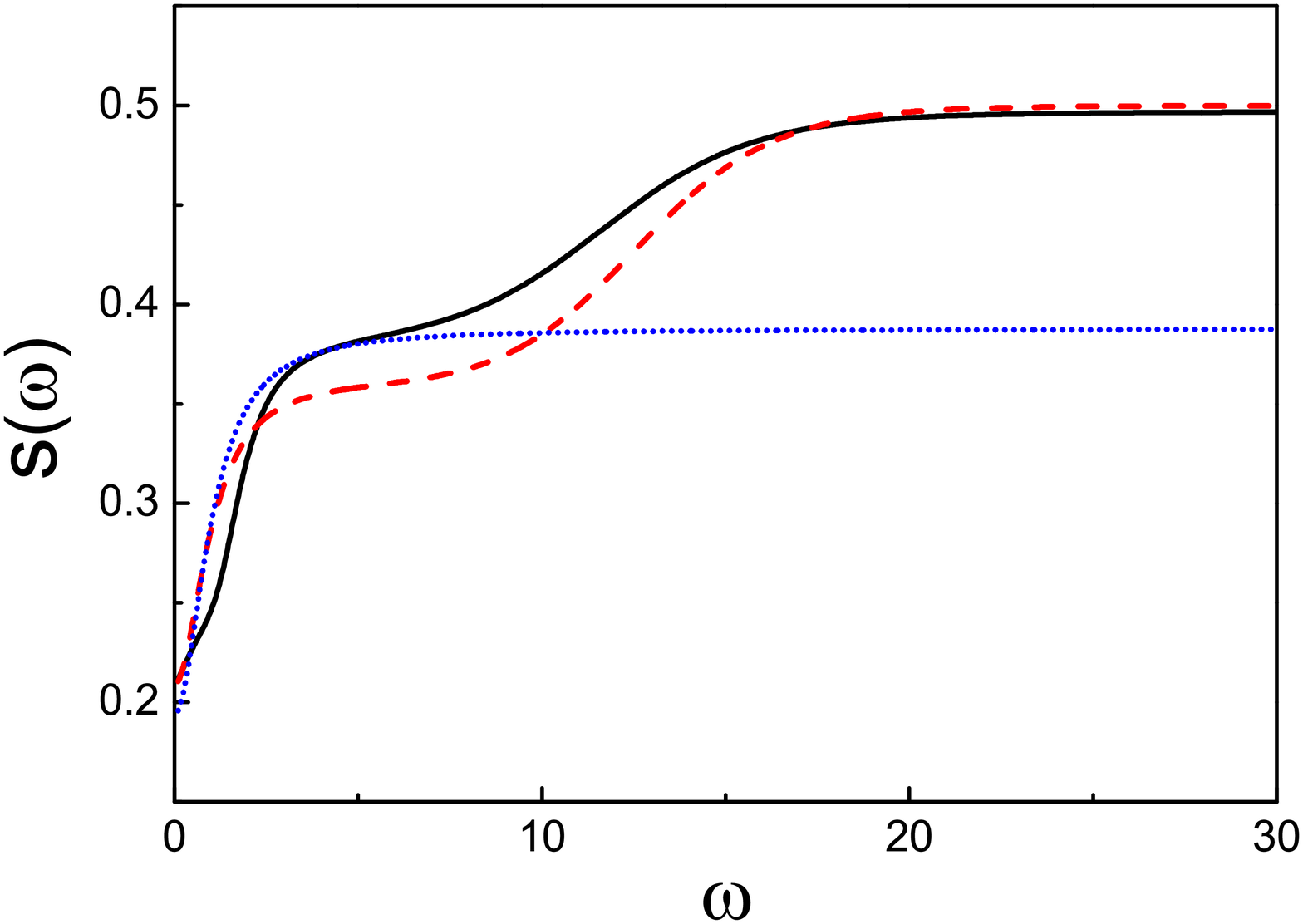}
\caption{
Shot noise spectrum through a single-level noninteracting quantum dot,
from the SCBA-ME (solid curve), the 2nd-nMKV-ME (dashed curve)
and the 2nd-MKV-ME (dotted curve), respectively.
Parameters: $\Gamma_L=\Gamma_R=0.5$,
$\mu_L=-\mu_R=7.5$,
$\epsilon_0=5$, $k_BT=2$ and $W=100$.  }
\end{figure}

\subsection{Coulomb-Blockade Quantum Dot }

This is the system described by \Eq{Ands-H}. Here we consider
first the noise spectrum in the Coulomb-Blockade (CB) regime,
while leaving the Kondo regime in next subsection.
The CB regime of single occupation is characterized by
$\epsilon_0+U>\mu_L>\epsilon_0>\mu_R$.
For the purpose of comparison, we quote
the result from the 2nd-MKV-ME \cite{Luo2007}:
\begin{align}\label{CBSW}
S(\omega) = 2\bar{I} \left[
\frac{4\Gamma^2_L+\Gamma^2_R+\omega^2}
{(2\Gamma_L+\Gamma_R)^2+\omega^2} \right].
\end{align}
In large bias limit, i.e., the Fermi levels being far from
$\epsilon_0$ and $\epsilon_0+U$, the steady state current
reads $\bar{I}=2\Gamma_L\Gamma_R/(2\Gamma_L+\Gamma_R)$.
However, in numerical simulation we account for the finite bias effect by
inserting the steady state current from the SCBA-ME approach into \Eq{CBSW}.
Notice that in obtaining \Eq{CBSW}
the double occupancy of the dot is excluded
because its energy is out of the bias window.
In the $n$-SCBA-ME treatment, however,
all the four basis states should be included.

In Fig.\ 3 we display the main result of the noise spectrum in the CB regime,
where a couple of non-Markovian resonance steps are revealed at frequencies
around $\omega_{\alpha 0}=|\mu_{\alpha}-\epsilon_0|$
and $\omega_{\alpha 1}=\epsilon_0+U-\mu_{\alpha}$.
We find that the resonance steps in high frequency regime
are enhanced by the Coulomb interaction, while the low frequency
spectrum has remarkable ``renormalization" effect compared to \Eq{CBSW}.
In addition to the result under the wide band limit (WBL),
in Fig.\ 3 we also show the bandwidth effect by two more curves.
We see that, for finite-bandwidth leads, the noise spectrum
diminishes at high frequency limit.
This is because the energy ($\omega$) absorption/emission
of detection restricts the channels for electron transfer
between the dots and leads.

\begin{figure}
\includegraphics[scale=0.28]{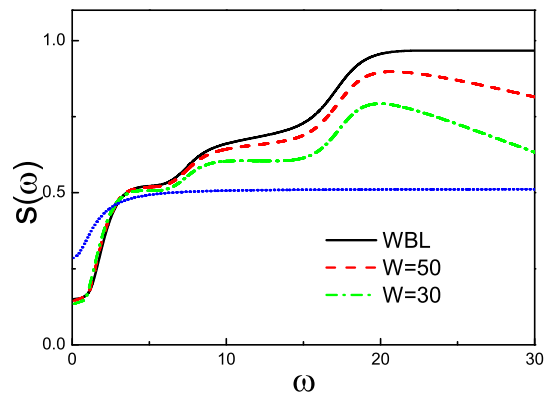}
\caption{
Shot noise spectrum through an interacting quantum dot
in a Coulomb blockade regime defined as
$\epsilon_0+U>\mu_L>\epsilon_0>\mu_R$.
Results in the wide band limit (WBL)
and for the finite bandwidths are shown
in comparison with the one from the 2nd-MKV-ME (dotted line).
Parameters: $\Gamma_L=\Gamma_R=0.5$, $\mu_L=-\mu_R=5$,
$\epsilon_{\uparrow}=\epsilon_{\downarrow}=\epsilon_0=2$,
$U=10$, and $k_BT=2$. We find staircases
appearing at $\omega=\epsilon_0-\mu_R=7$
and $\epsilon_0+U-\mu_R=17$.  }
\end{figure}


\subsection{Nonequilibrium Kondo Dot}

The nonequilibroum Kondo system, with the Anderson impurity model
realized by transport through a small quantum dot,
has been attracted intensive attention in the past two decades
\cite{Gor98,Kou98,Gla04,Ng88,Her91,MW92,MW9193,Ra94,Mar06,Gor08,Yan12}.
Compared to the {\it equilibrium} Kondo effect,
the {\it nonequilibrium} is characterized by
a finite chemical potential difference of the two leads.
As a result, the peak of the density of states (spectral function)
splits into two peaks pinned at each chemical potential.
The two peak structure is difficult to probe directly,
by the usual dc measurements.
Nevertheless, the shot noise can be a promising quantity
to reveal the nonequilibrium Kondo effect,
although much less is known about it.
We notice that results on low-frequency
noise measurements have only appeared very recently \cite{De09,Hei08},
while so far there are not yet reports
on the finite-frequency (FF) noise measurements.
A couple of theoretical studies \cite{Ng97,Her98,Kon07,Moc11}, however,
revealed diverse signatures (Kondo anomalies) in the FF noise spectra,
such as an ``upturn" \cite{Ng97} or a spectral ``dip" \cite{Moc11}
appeared at frequencies $\pm eV/\hbar$ ($V$ is the bias voltage),
as well as the Kondo singularity (discontinuous slope)
at frequencies $\pm 2eV/\hbar$ in Ref.\ \cite{Her98},
or at $\pm eV/2\hbar$ in Ref.\ \cite{Moc11}.
Also, it was pointed out in Ref.\ \cite{Her98} that
the minimum (dip) developed at $\pm eV/\hbar$
is not relevant to the Kondo effect,
since in the noninteracting case the noise
has similar discontinuous slope at $\pm eV/\hbar$ as well.

The system Hamiltonian is still \Eq{Ands-H}, which corresponds
to the well known Anderson impurity model.
Following the solving protocol outlined at the end of Sec. III
together with the results in Sec. II (D), we obtain the
noise spectrum in the Kondo regime as shown in Fig.\ 4.
Remarkably, we notice a profound {\it dip} behavior (Kondo signature)
in the noise spectrum at frequencies $\omega=\pm V/2$,
as particularly demonstrated by a couple of voltages.
We attribute this behavior to the emergence of the
Kondo resonance levels (KRLs) at the Fermi surfaces,
i.e., at $\mu_L=V/2$ and $\mu_R=-V/2$.
In steady state transport, it is well known that
the KRLs are clearly reflected in the spectral function.
In terms of the master equation,
the KRLs structure is hidden in the self-energy terms,
which characterize the tunneling process
and define the transport current.
Similarly, the noise spectrum is essentially affected,
particularly in the Kondo regime,
by the self-energy process in frequency domain
based on the same master equation.
This explains the emergence of the spectral dip appearing
at the same KRLs (i.e., at $\omega=\pm V/2$).

Alternatively, as a heuristic picture, one may imagine to include
the KRLs as basis states in propagating $\rho(t)$,
which is implied in the current correlation function.
In usual case, when the level spacing is larger than its broadening,
the diagonal elements of the density matrix decouple to
the evolution of the off-diagonal elements.
However, in the Kondo system, the diagonal and off-diagonal elements
are coupled to each other, through the complicated self-energy processes.
This feature would bring the {\it coherence evolution}
described by the off-diagonal elements, with characteristic
energies of the KRLs and their difference, into the diagonal elements
which contribute directly to the the second current measurement
in the correlation function $\la I(t) I(0) \ra$.
Then, one may expect three coherence energies, $\pm V/2$ and $V$,
to participate in the noise spectrum.
Indeed, the dip emerged in Fig.\ 4 reveals the {\it coherence} induced
oscillation at the frequencies $\pm V/2$,
while the other one at the higher frequency $V$
(observed in Ref.\ \cite{Moc11} in the case of infinite $U$)
is smeared in our finite $U$ system by the rising noise with frequency.

\begin{figure}
\includegraphics[scale=0.28]{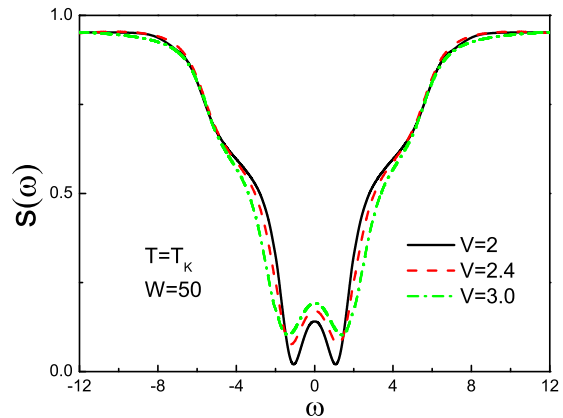}
\caption{
Shot noise spectrum in the Kondo regime,
for several bias voltages ($\mu_L=-\mu_R=V/2$).
Parameters: $\Gamma_L=\Gamma_R=\Gamma_0=0.5$,
$\epsilon_{\uparrow}=\epsilon_{\downarrow}=\epsilon_0=-2$, and $U=6$.
The Kondo temperature is determined by
$T_K=\frac{U}{2\pi}\sqrt{\frac{-2U\Gamma_0}{\epsilon_0(U+\epsilon_0)}}
\exp[\frac{\pi\epsilon_0(U+\epsilon_0)}{2U\Gamma_0}]$,
being thus $T_{K}=0.144$ for the given parameters.    }
\end{figure}

\section{Summary}

To summarize, in this work we propose
a particle-number-resolved transport master equation
under self-consistent Born approximation.
The most advantage of this approach is its efficiency
in the study of shot noise and a potential application
in counting statistics.
We have demonstrated this new approach by several examples,
including particularly the nonequilibrium Kondo system.
The obtained results are completely
beyond the scope of the Born-Markov master equation,
revealing such as staircase behavior and the profound nonequilibrium
Kondo signature in the shot noise spectrum.
The validity of the proposed approach is also supported by
the evidence in steady state, where this approach can recover not only
the exact result of noninteracting transport under arbitrary voltages,
but also the challenging nonequilibrium Kondo effect.

\vspace{0.5cm}
{\it Acknowledgements.}---
This work was supported by the NNSF of China,
the Major State Basic Research Project of China
under grants 2011CB808502 \& 2012CB932704,
and the Fundamental Research Funds for the Central Universities of China.
J.J. was also supported by the Program for Excellent Young Teachers
in Hangzhou Normal University and by the NSFC under No.11274085.




\end{document}